\documentclass[aps, 10pt, a4paper, floatfix, rmp, twocolumn, showkeys, superscriptaddress]{revtex4-2}

\usepackage[utf8]{inputenc}
\usepackage{booktabs}
\usepackage{graphicx}
\usepackage{lmodern}
\usepackage{mathtools}
\usepackage{amsfonts, amssymb}
\usepackage{calrsfs}
\usepackage{silence}
\usepackage{xcolor}
\usepackage[most]{tcolorbox}

\usepackage[hidelinks]{hyperref}

\setcitestyle{numbers,square,comma}

\makeatletter
\def\NAT@sort{\@ne}
\def\NAT@cmprs{\@ne}
\makeatother

\NewTColorBox[auto counter]{widegraybox}{O{} O{}}{
    float*,
    floatplacement={!t},
    width=\textwidth,
    colback=gray!10,
    colframe=black!65,
    boxrule=0.5pt,
    arc=2pt,
    before skip=1em,
    after skip=1em,
    title={Box~\thetcbcounter:~#1},
    fonttitle=\strut\bfseries,
    coltitle=white,
    #2,
}

\IfFileExists{etoolbox.sty}{%
    \ifdef{\iflatexml}{
        \newcommand{\printEndNotes}{}%
    }{
        \usepackage{etoolbox}
        \newcounter{myendnotecount}
        \newcommand{\myEndNotes}{}
        \newcommand{\addEndNote}[1]{%
            \stepcounter{myendnotecount}%
            \hyperlink{Note.\themyendnotecount}{%
                \textsuperscript{\themyendnotecount}%
                \hypertarget{NoteReference.\themyendnotecount}%
            }%
            \xappto\myEndNotes{%
                \begingroup%
                    \clubpenalties 5 1000 1000 1000 1000 100%
                    \widowpenalties 5 1000 1000 1000 1000 100%
                    \noexpand\noindent%
                    \noexpand\hyperlink{NoteReference.\themyendnotecount}{%
                        \noexpand\textsuperscript{\themyendnotecount}%
                        \noexpand\hypertarget{Note.\themyendnotecount}{}%
                    }%
                    \noexpand\unskip\noexpand\space\noexpand\ignorespaces%
                    \unexpanded{##1}%
                    \noexpand\par\noexpand\medskip%
                \endgroup%
            }%
        }
        \renewcommand{\footnote}[1]{\addEndNote{##1}}
        \newcommand{\printEndNotes}{%
            \section*{Notes}%
            \myEndNotes%
        }%
    }%
}{%
    \typeout{Warning: etoolbox not found. Endnote system disabled.}
    \newcommand{\printEndNotes}{}
}

\begin{document}

\title{Cognition spaces: natural, artificial, and hybrid}

\providecommand{\CSL}{Complex Systems Lab, Universitat Pompeu Fabra, Dr. Aiguader 88, 08003 Barcelona.}
\providecommand{\ICREA}{Instituci{\'o} Catalana de la Recerca i Estudis Avançats (ICREA), Pg. Lluís Companys 23, 08010 Barcelona.}
\providecommand{\IBE}{Institut de Biologia Evolutiva, CSIC-UPF, Pg. Mar{\'i}tim de la Barceloneta 37, 08003 Barcelona.}
\providecommand{\SFI}{Santa Fe Institute, 1399 Hyde Park Road, Santa Fe, NM 87501, United States.}
\providecommand{\MELIS}{Universitat Pompeu Fabra, Medicine and Life Sciences Department (MELIS), Barcelona.}
\providecommand{\SCOMPUT}{School of Computing, Australian National University, ACT, Australia.}
\providecommand{\ISEM}{ISEM, Université de Montpellier, CNRS, IRD 34095 Montpellier, France}
\providecommand{\TUFTS}{Allen Discovery Center, Tufts University, Medford, MA, United States.}
\providecommand{\WYSS}{Wyss Institute for Biologically Inspired Engineering, Harvard University, Boston, MA, United States.}

\author{Ricard Sol\'{e}}
\email[Corresponding author, ]{ricard.sole@upf.edu}
\affiliation{\CSL}
\affiliation{\ICREA}
\affiliation{\IBE}
\affiliation{\SFI}

\author{Luis F Seoane}
\affiliation{\IBE}

\author{Jordi Pla-Mauri}
\affiliation{\CSL}
\affiliation{\IBE}

\author{Michael Timothy Bennett}
\affiliation{\SCOMPUT}

\author{Michael E. Hochberg}
\affiliation{\ISEM}
\affiliation{\SFI}

\author{Michael Levin}
\affiliation{\TUFTS}
\affiliation{\WYSS}

\vspace{0.4 cm}
\begin{abstract}
\vspace{0.2 cm}
Cognitive processes are realized across an extraordinary range of natural, artificial, and hybrid systems, yet there is no unified framework for comparing their forms, limits, and unrealized possibilities. Here, we propose a cognition space approach that replaces narrow, substrate-dependent definitions with a comparative representation based on organizational and informational dimensions. Within this framework, cognition is treated as a graded capacity to sense, process, and act upon information, allowing systems as diverse as cells, brains, artificial agents, and human--AI collectives to be analyzed within a common conceptual landscape. We introduce and examine three cognition spaces---basal aneural, neural, and human--AI hybrid---and show that their occupation is highly uneven, with clusters of realized systems separated by large unoccupied regions. We argue that these voids are not accidental but reflect evolutionary contingencies, physical constraints, and design limitations. By focusing on the structure of cognition spaces rather than on categorical definitions, this approach clarifies the diversity of existing cognitive systems and highlights hybrid cognition as a promising frontier for exploring novel forms of complexity beyond those produced by biological evolution.
\end{abstract}


\maketitle

\begin{quote}
    \raggedleft\footnotesize
    ``The nervous system, which probably first evolved as a device to coordinate the behaviour of various cells in multicellular organisms (\dots) ultimately became able to invent the future''\par
    \smallskip
    \textit{The Possible and The Actual}\\
    François Jacob\par
\end{quote}

\section{Introduction}

What kinds of cognitions exist? \cite{Sloman1984Structure, Yampolskiy2015Space, ball2022book} What kinds of minds could exist but have never evolved? What kinds of constraints might limit the possible when dealing with minds? \cite{roth2015convergent, powell2017convergent, sole2024fundamental}
What forms of cognitive complexity might be engineered beyond those realized by living systems?
Cognition, when broadly defined as the capacity of a system to sense, process, and respond to information, is not restricted to humans or even to neural substrates \cite{Baluka2016, gunawardena2022learning, Kaygisiz2025Molecular, koseska2017cell, Levin2022Technological, Lyon2005Biogenic, Lyon2019MinimalCognition, Rouleau2023Multiple}. Evidence across biology,  physics, and the synthetic sciences indicates that cognitive processes can emerge within a wide range of material and organizational substrates, from single cells \cite{Adamatzky2010, koseska2017cell, gunawardena2022learning} and multicellular organisms \cite{Sole2019}, to swarms \cite{bonabeau1997self}, artificial neural networks \cite{Edelman2005} and distributed sociotechnical systems \cite{Hutchins1995, Hollan2000}.
Despite this diversity, however, contemporary accounts of cognition remain shaped by 
disciplinary boundaries and, in some cases, anthropocentric assumptions about what qualifies as a ``thinking'' or ``learning'' system.
A recurring challenge in the study of cognition and related notions such as intelligence or consciousness is the lack of consensus on a definition that is sufficiently general to encompass the full spectrum of possible cognitive processes and systems.

\begin{figure*}[t]
    \centering
    \includegraphics[width=0.8\linewidth]{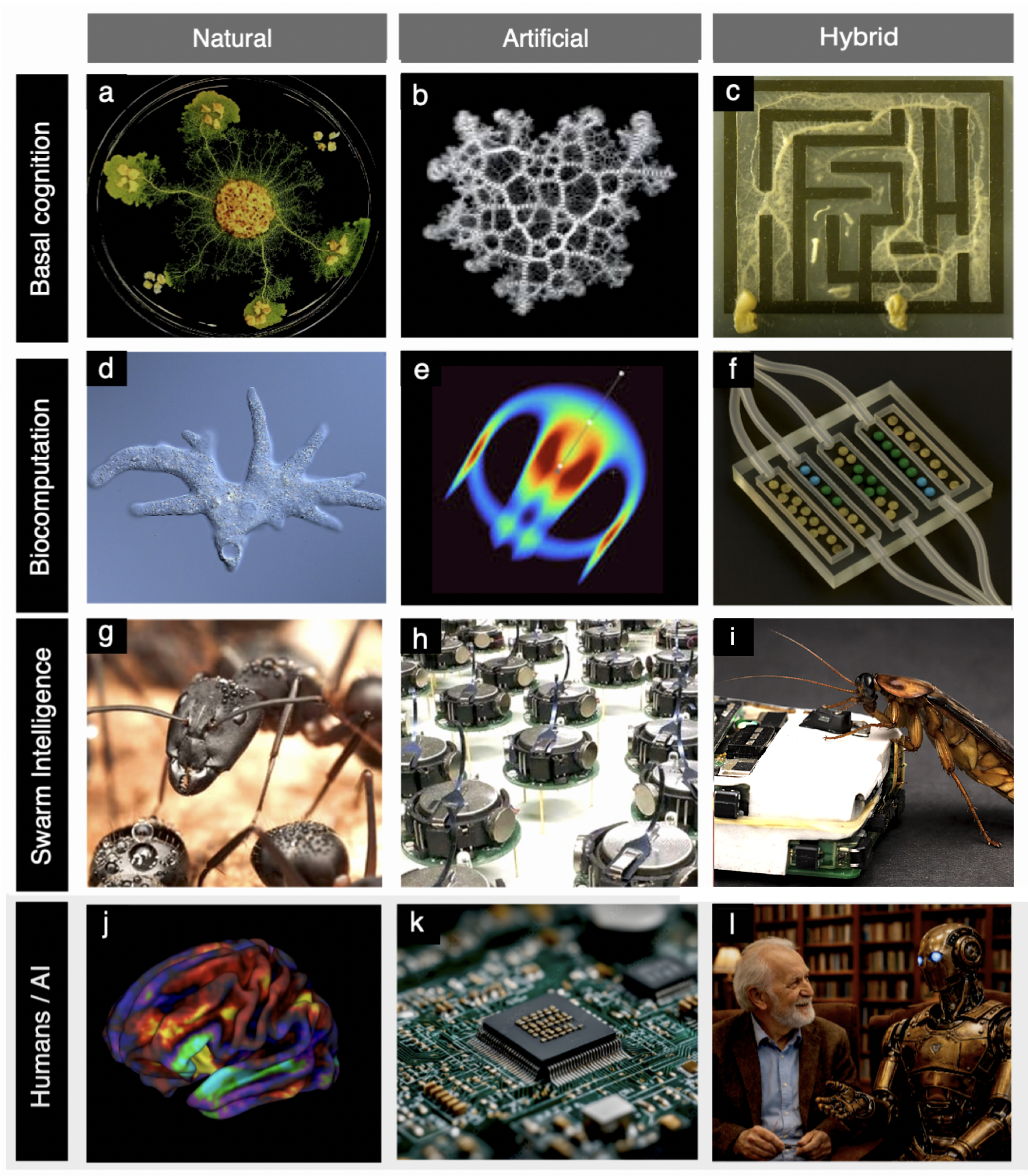}
    \caption{Natural, artificial, and hybrid cognition across organizational scales.
Representative systems are organized along three columns---natural, artificial, and hybrid---and levels of organizational complexity: basal cognition, 
biocomputation, swarm intelligence, ecosystems, and humans/AI. Examples in (a--c) illustrate basal cognitive and path-finding processes in homogeneous media (a), simulated particle swarms (b), and hybrid settings (c) as defined by human-designed graphs. In (d--f), different systems performing computations are shown: (d) an amoeba, (e) Lenia, and (f) a microfluidic device with spatially segregated cell strains. In (g--i), we display collective intelligence examples from social insects (g), robotic swarms (h), and biohybrid insect--robot systems (i). Finally, when dealing with higher cognition and AI (j--l), highlight human brains (j), neuromorphic computers (k), and human--AI dyads (l).}
    \label{fig:cognition_scales}
\end{figure*}

Debates about the nature of cognition have long revolved around whether it should be understood as a graded, continuous property of natural systems or as a discrete kind that appears only once certain organizational or functional thresholds are crossed \cite{Dennett1996Kinds, Haig2020From, Levin2022Technological}.
Continuum-based views emphasize that many of the mechanisms underlying known examples of cognition, such as bioelectrical networks~\cite{Levin2023Bioelectric, martinez2019metabolic}, and algorithms, such as information processing, feedback control, learning, adaptive regulation, and active inference~\cite{Friston2013Life, Kirchhoff2018Markov, Pezzulo2018Hierarchical, Rubin2020FutureClimates, Solms2019ConsciousnessFreeEnergy, Solms2024StrangeParticle, Parr2022ActiveInference} are already present, in rudimentary forms, across a wide range of biological and even non-biological systems at many scales~\cite{sperl1999hebbian, Hanczyc2014Droplets, Cejkova2017Droplets, Biswas2021, Biswas2022}.
In this sense, it has been speculated that differences between cellular chemotaxis, animal perception, and reasoning arise primarily from changes in degree and organization, rather than from categorical differences in kind \cite{Bentley2014Endothelial, Boisseau2016Habituation, Boussard2019Memory, eckert2024biochemically, Kramar2021Encoding, kukushkin2024massed, Levin2023Bioelectric, mcmillen2024collective, Spencer2003Potentiation, Vogel2016TransferBehaviour}.

By contrast, a discontinuous view of cognition holds that cognitive systems belong to a finite number of discrete classes. In this sense, cognition requires specific architectures or representational capacities. 
These are often tied to qualitative features such as movement, nervous systems, symbolic manipulation, or intentional states. 
Systems that lack these features may still display complex or adaptive behavior. 
However, they are not genuinely cognitive. 
Proponents of this view warn that overly permissive definitions risk trivializing cognition. 
Such definitions may extend cognition to any self-organizing or responsive system. The 
emerging field of diverse intelligence treats this as an experimental question. 
It is not primarily a linguistic or philosophical one.
The goal is not to preserve the inherited conceptual categories. 
Instead, tools from cybernetics and the cognitive and behavioral sciences are tested across novel substrates. 
The aim is to empirically determine when these tools offer practical advantages. 
It is also to determine when they do not apply in a useful way~\cite{Levin2022Technological}. 
Nevertheless, both continuum and discontinuous perspectives rely on non-trivial theoretical assumptions. 
These assumptions concern which organizational features are cognitively relevant. 
The tension between these views reflects a deeper methodological choice. 
Cognition may be defined by minimal functional criteria. 
Such criteria allow smooth transitions across systems. 
Alternatively, cognition may be defined by stronger conditions. 
These seek to classify nature into distinct kinds.

Is there a way to accommodate the diversity of cognitive systems while still providing principled answers to the questions posed above? Within biology, closely related challenges arise across systems and scales. Consider, for example, the long-standing problem of defining life. Despite sustained efforts across disciplines, no single definition has achieved broad consensus, suggesting that an alternative conceptual strategy may be required \cite{Bender2025What}. An influential approach is to take diversity itself as the starting point: rather than seeking necessary and sufficient conditions, one examines a wide range of case studies and situates them within a space defined by a small set of relevant dimensions, such as information processing, structural organization, and metabolic autonomy. When living (and in some cases non-living) systems are mapped into such a space, a striking pattern often emerges, characterized by highly uneven occupation and extensive voids \cite{raup1966geometric, Olle2016Morphospace, sole2025origins, dera2026mapping}. Both features are highly informative. Regions of dense occupation reveal recurrent combinations of traits and provide clues about how different forms of complexity have co-evolved, while unoccupied regions may signal effectively inaccessible domains, i.e. configurations that are evolutionarily inaccessible, physically or chemically constrained, or dynamically unstable. The resulting picture is both rich and interpretable: a morphospace that does not merely catalog diversity, but helps explain it.

In this paper, we adopt this conceptual framework to introduce and analyse three spaces of cognition spanning natural, artificial, and hybrid systems. Figure~\ref{fig:cognition_scales} surveys a set of case studies encompassing natural, artificial, and hybrid systems that combine living and non-living components. Our aims are threefold. First, we address the problem of cognitive complexity by situating diverse case studies within a set of relevant spaces, thereby shifting attention away from potentially ambiguous definitions of cognition. On this view, the structure of the space itself and the patterns of its occupation constitute an appropriate response to key questions about the nature and diversity of cognition. Second, we interpret the voids separating clusters within these spaces in terms of design principles, engineering challenges faced by artificial cognitive systems, and evolutionary or physical constraints. Finally, across all three spaces, we undertake a comparative analysis of natural and artificial systems while explicitly examining hybrid forms of cognition that integrate both. Specifically, we define and analyse three cognition spaces, namely:
\begin{itemize}
    \item \textbf{Basal, aneural cognition space:} single cells, simple multicellular organisms, xenobots, and organoids can exhibit minimal yet rich forms of memory, learning, and perception--action loops, offering minimal models of living information processing.

    \item \textbf{Neural cognition space:} multicellular systems equipped with neurons and brains, along with artificial agents. Synthetic biology and microbiome engineering reveal how designed genetic circuits can reshape behavioral attractors, introducing new modes of learning, coordination, and decision-making.

    \item \textbf{Human--AI hybrid cognition space:} interactions between AI systems (e.g., large language models) and human users exemplify reciprocal learning processes. Several levels of pairwise complexities arise, including emergent classes of hybrid cognitive ecosystems---what we may call the \emph{humanbot}, reflecting tightly-coupled but still distinctive agencies.
\end{itemize}

As we shall see, defining hybrid systems is far from trivial. Strictly speaking, hybridity no longer resides solely in physical attachment or
material fusion. Systems such as Xenobots are not merely assemblies of living cells, but extended networks that include human experimenters, 
AI design tools, and robotic platforms that actively explore biological possibility spaces. These components interact through functional and 
informational interfaces, forming an integrated causal system. As a result, the boundary of the “system” becomes ambiguous: it is unclear 
where the organism ends, and its cognitive--technological scaffold begins. Hybrid systems are therefore increasingly characterized by 
distributed function and collective agency rather than by clear material or ontological criteria. By constructing cognition spaces, we aim to 
sharpen the distinction between qualitative classes of cognition, compare their complexities, and identify potential analogies and qualitative 
discontinuities among them.

\begin{figure*}[t]
    \centering
    \includegraphics[width=0.95\linewidth]{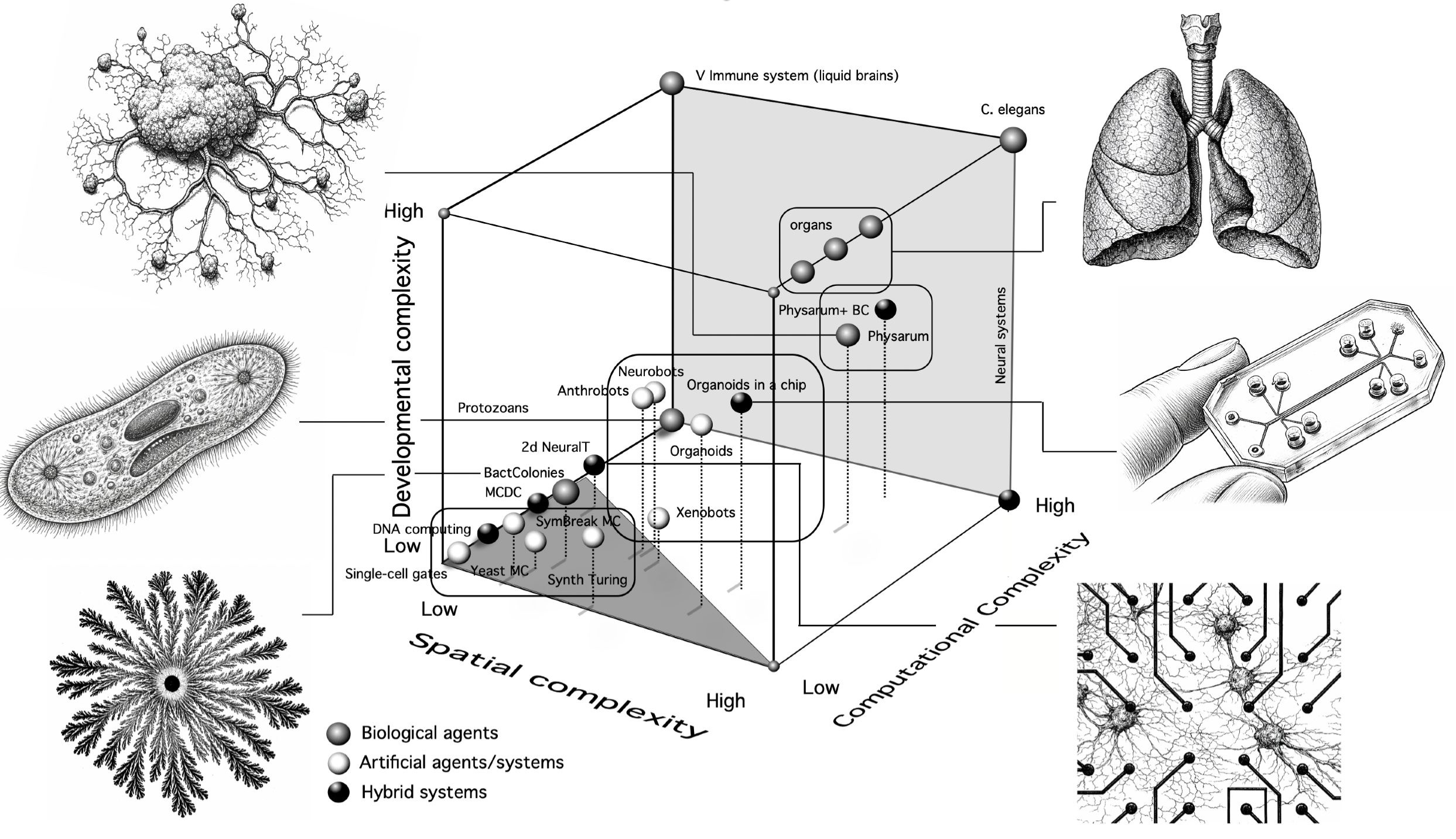}
    \caption{{\bf A Morphospace of basal cognition.}  
    A space of living (or hybrid). Here, biological, artificial, and hybrid systems are positioned within a three-dimensional space defined by spatial complexity, computational complexity, and developmental complexity. These axes capture (in an aggregated manner) the diversity and complexity of decision making, multicellular organization, and the role played by development. The diagram spans systems from single cells, bacterial colonies, and protozoa to multicellular organisms, organs, and engineered living systems such as organoids, xenobots, and neurally interfaced constructs. The dark shaded region at the bottom left highlights regimes of basal cognition based on synthetic systems. Illustrative sketches anchor representative systems along each axis, emphasizing the diversity of natural and engineered forms of minimal cognition and the presence of large voids.}
    \label{fig:basal_cognition} 
\end{figure*}

\section{Basal cognition space}

Our first space of cognition comprises systems that lack neural elements and exemplify basal cognition (Fig.~\ref{fig:basal_cognition}), understood as minimal cognitive capacities arising from embodied, self-organized dynamics rather than nervous systems. It encompasses processes such as sensing, information integration, adaptive responses, and learning-like plasticity. These processes occur in single cells, microbial collectives, and simple multicellular systems~\cite{Lyon2005Biogenic, Baluka2016, koseska2017cell, Levin2021, gershman2021reconsidering, levin2023darwin}. Our morphospace includes a diverse range of biological organizations, from bacteria (single cells, cell cultures, and engineered 
strains) to free-living eukaryotic cells (protozoans) and the giant single-celled slime mold. Alongside these natural 
systems, we include a growing class of multicellular synthetic constructs comprising two distinct categories of artificial living systems: (a) living bots (xenobots, anthrobots, and neurobots), built from animal or human cells, whose collective, goal-directed behavior arises from morphology and self-organization, and (b) organoids, that is, 
selforganized multicellular structures that recapitulate key features of organs. To define the boundaries of this space, we also include natural organs, the immune system, neurons on a chip \cite{eckmann2007physics}, and a simple neural organism (the nematode {\it C. elegans}). The axes are intended to capture spatial complexity, computational complexity, and the role of development in the generative 
processes underlying each system. The systems are positioned relative to each other; the space is not metric.

Where does cognition come from in the aneural world? How complex must the simplest organism that displays cognition be? Is a certain level of information processing necessary for the survival of its predecessors?
Some key components of cognition, such as memory, are even possible in small molecular networks~\cite{Biswas2022, Biswas2021, Csermely2020Learning, Keresztes2025Cancer, Veres2024CellularForgetting} (not requiring a cell, never mind a brain). Gene and signaling networks, as well as immune networks, are well known to be describable in formal terms as a class of neural-like networks \cite{kauffman1992origins}. In some cases, information processing is grounded in morphology. In such cases, the result of a computational process is allowed by the exploratory behavior of their internal networks \cite{kirschner1997cells}. {\it Physarum polycephalum} (Fig.~\ref{fig:cognition_scales}a) exhibits morphological computation on the organismal scale, where its adaptive network of tubes physically encodes optimization and memory through growth and flow dynamics \cite{Adamatzky2010}. In contrast, an amoeba (Fig.~\ref{fig:cognition_scales}d) relies on cytoskeletal plasticity on the intracellular scale, where rapid actin--myosin reorganization enables flexible sensing, movement, and decision-making, embedding computation directly in cellular mechanics \cite{nalbant2018exploratory, fletcher2010cell, hohmann2019cytoskeleton}.

\begin{widegraybox}[Physarum on a graph as hybrid cognition][label=box:physarum]
    The slime mold \textit{Physarum polycephalum} provides a canonical example of hybrid cognition, in which an adaptive biological system exploits an external, human-imposed structure to solve problems of path finding and network optimisation. In artificial settings, \textit{Physarum} is often constrained to grow on a predefined \emph{graph}, that is, a discrete spatial structure $G=(V, E)$ consisting of a set of nodes $V$ connected by edges $E$. In practice, nodes correspond to nutrient sources or sinks, while edges represent permissible pathways for transport, defined by channels, substrates, or patterned environments designed by an experimenter.
    
    The graph, therefore, acts as an external boundary condition that encodes topological and metric information not generated by the organism itself. Within this constrained environment, the dynamics of transport and growth can be expressed through a variational principle. The \textit{Physarum} Lagrangian $\mathcal{L}_{\phi}=\mathcal{L}_{\phi}(Q,\lambda;D)$ is defined on the graph $G$ as
    \begin{equation}
        \mathcal{L}_{\phi}
        = \frac{1}{2} \sum_{(i, j)\in E} \frac{L_{ij}}{D_{ij}}\, Q_{ij}^2
        + \sum_{i\in V} \lambda_i \left( S_i - \sum_{\mathclap{j:(i, j)\in E}} Q_{ij} \right),
        \label{eq:Lagrange}
    \end{equation}
    where $Q_{ij}$ denotes the flux along edge $(i, j)$, $L_{ij}$ its length, $D_{ij}$ its adaptive conductivity, and $S_i$ the prescribed source or sink strength at node $i$. The Lagrange multipliers $\lambda_i$ enforce local conservation of flow at each node.
    
    The first term in $\mathcal{L}_{\phi}$ represents the energetic cost of transport along the graph. Its quadratic dependence on the flux $Q_{ij}$ and inverse dependence on the conductivity $D_{ij}$ encode dissipation: high flux through long or weakly conducting edges is costly, while reinforced edges (large $D_{ij}$) reduce dissipation. The second term imposes the topological constraint that inflow and outflow balance at each node, except for externally specified sources and sinks. \emph{Formally, the system is equivalent to a resistive electrical network in which currents distribute to minimise total power dissipation.} Minimisation of $\mathcal{L}_{\phi}$ yields flux configurations that balance transport efficiency against network structure, resulting in the selection of short, well-connected paths. Crucially, the optimisation encoded by $\mathcal{L}_{\phi}$ is not ``computed'' internally by the organism in any representational sense. Instead, it emerges from the coupling between the mold's local growth and reinforcement dynamics and the externally imposed graph. In this setting, the graph functions as a human-designed cognitive scaffold that restricts and shapes the space of viable solutions, while \textit{Physarum}'s embodied dynamics explore and exploit that space. Path finding in constrained \textit{Physarum} systems thus exemplifies hybrid cognition: the solution arises from the joint action of biological adaptation and engineered structure, rather than from either component alone.
\end{widegraybox}

Hybrids have been obtained in different ways, and a few are shown in Fig.~\ref{fig:basal_cognition} as black spheres. 
An important class of hybrid agencies is the result of the blend of a living system (distributed in a given spatial context) with an externally imposed set of boundary conditions (BCs). These BCs provide a set of physical constraints that either allow us to exploit the intrinsic biology of the organism or trigger some physicochemical process that allows the system to achieve higher complexity states. 
The slime mold \textit{Physarum polycephalum} provides a canonical illustration of basal cognition and its extension using hybrid designs. When not constrained, its plasmodial network expands irregularly, following local gradients of chemoattractants. However, when placed within a structured environment, such as a maze (Fig.~\ref{fig:cognition_scales}c), a graph of nutrient nodes, or a constrained microchannel, it exhibits emergent problem-solving behaviors. The oscillatory contractile flows of the mold, normally diffusive and disordered, become spatially organized to cover and connect the nutrient sites through minimal transport paths. The maze walls or the nutrient graph, though inert, restrict the configuration space of the plasmodium and thus induce a global optimization process that appears to be aimed. The imposed geometry does not provide information about the optimal solution, but filters the spontaneous dynamics of the organism so that efficient patterns are preferentially stabilized. In this way, \textit{Physarum} exhibits computationally interpretable behavior through interaction with an \emph{inert constraint field}. This can be conceptualized using a variational principle\footnote{
Let ${\bf x} = (x_1, \dots, x_n)$.
The method of Lagrange multipliers provides a necessary condition for the extrema of a function 
$f({\bf x})$ subject to $M$ constraints $g_k({\bf x})=0$, with $k = 1, \ldots, M$.
This is done by building the augmented function or {\it Lagrangian}
\[
\mathcal{L}({\bf x},\lambda_1,\ldots,\lambda_M)
= f({\bf x}) - \sum_{k=1}^M \lambda_k g_k({\bf x}),
\] where $\lambda_k$ are Lagrange multipliers.
Stationary points are determined by requiring
\begin{align*}
    \frac{\partial \mathcal{L}}{\partial x_i} &= 0, & \text{for all } i &= 1, \dots, n,
\end{align*}
together with the original constraints, \( g_k({\bf x}) = 0 \), for all \( k = 1, \dots, M \).
\\
This procedure transforms the original constrained optimization problem into a solvable system of equations.
}, as summarized in Box~\ref{box:physarum}. BCs have also been used to construct synthetic combinatorial circuits based on libraries of engineered cells arranged through predefined spatial segregation (Fig.~\ref{fig:cognition_scales}f; MCDC in Fig.~\ref{fig:basal_cognition}). This hybrid system mimics the spatial segregation of cell types found in an organism, yet it does not adhere to the standard design logic of either electronic or biological circuits\footnote{
A logic multicellular circuit implements a Boolean function
$f:\{0, 1\}^N \rightarrow \{0, 1\}$ mapping $N$ molecular inputs
$\{x_j\}$ to an output.
In canonical (sum-of-products) form, the function can be written as
\[
f = \sum_{i=1}^{M} \left[ \prod_{j=1}^{N} \phi_{ij}(x_j) \right],
\]
where $\Sigma$ and $\Pi$ denote OR and AND operators, respectively,
$\phi_{ij}(x_j)\in\{x_j,\neg x_j\}$,
and $M\leq 2^{N-1}$.
Although universal, this representation requires extensive wiring
and genetic resources in biological implementations.
By applying double negation and De Morgan's laws, the Boolean function
can be reformulated as an OR of inverted OR-modules,
\[
f = \sum_{i=1}^{M} \psi_i,
\qquad
\psi_i = \overline{\sum_{j=1}^{N} \theta_{ij}(x_j)},
\]
with $\theta_{ij}(x_j)\in\{x_j,\neg x_j\}$.
This Inverted Logic Formulation (ILF) enables distributed computation
across spatially segregated cell consortia, substantially reducing
wiring requirements and favoring scalable biological circuit design.
} \cite{sole2013expanding, Macia2014How, Macia2016LogicConsortia}.

An analogous phenomenon occurs in multicellular, tissue-like systems such as organoids \cite{Olle2016Morphospace, garreta2019fine, takebe2019organoids, garreta2021rethinking, hofer2021engineering, zhao2022organoids}. When grown in isotropic environments, organoids typically remain amorphous and weakly organized. In contrast, culture within structured contexts---most notably microfluidic chips providing controlled flows, geometries, and BCs---induces new morphogenetic regimes \cite{takebe2017synergistic}. For example, cerebral organoids under perfusion develop cortical-like layered structures \cite{qian2019brain}, while intestinal organoids \cite{almeqdadi2019gut} and kidney organoids \cite{nishinakamura2019human} exposed to laminar flow exhibit spatially differentiated domains aligned with external gradients. In these hybrid systems, the artificial environment supplies BCs that break symmetry and reshape feedbacks between gene expression, mechanics, and transport, enabling higher levels of morphological and functional complexity not genetically pre-specified but induced by the external context. For comparison, we also included a hybrid system based on neuronal cultures on a chip \cite{soriano2023neuronal} (2d NeuralT in Fig.~\ref{fig:basal_cognition}).

Hybrots provide a distinct class of hybrid agents exhibiting basal cognition: biohybrid systems that couple living components of plant, animal, or fungal origins with technological interfaces and silicon-embedded algorithms \cite{Ando2020InsectMachine, Bakkum2007MEART, Clawson2022Endless, CohenKarni2012Smartest, Ding2018Cellular, Giselbrecht2013Chemistry, Li2016BrainComputer, Li2019Cyborg, Mehrali2018Blending, Orive2019CyborgScience, Aaser2017Towards, PioLopez2021Biocyborg, Potter2004Hybrots, Potter2003NeuronInterfaces, Saha2020Explosive, Tsuda2009PhiBot, Warwick2014CyborgRevolution, Warwick2011Experiments}. In these systems, biological substrates provide rich adaptive dynamics, whereas artificial components impose sensing, actuation, and control, forming closed perception-action loops. As a result, cognitive behavior emerges from the interaction between living matter and engineered structures, rather than being localized in either component alone.

\begin{figure*}[t]
    \centering
    \includegraphics[width=0.9\linewidth]{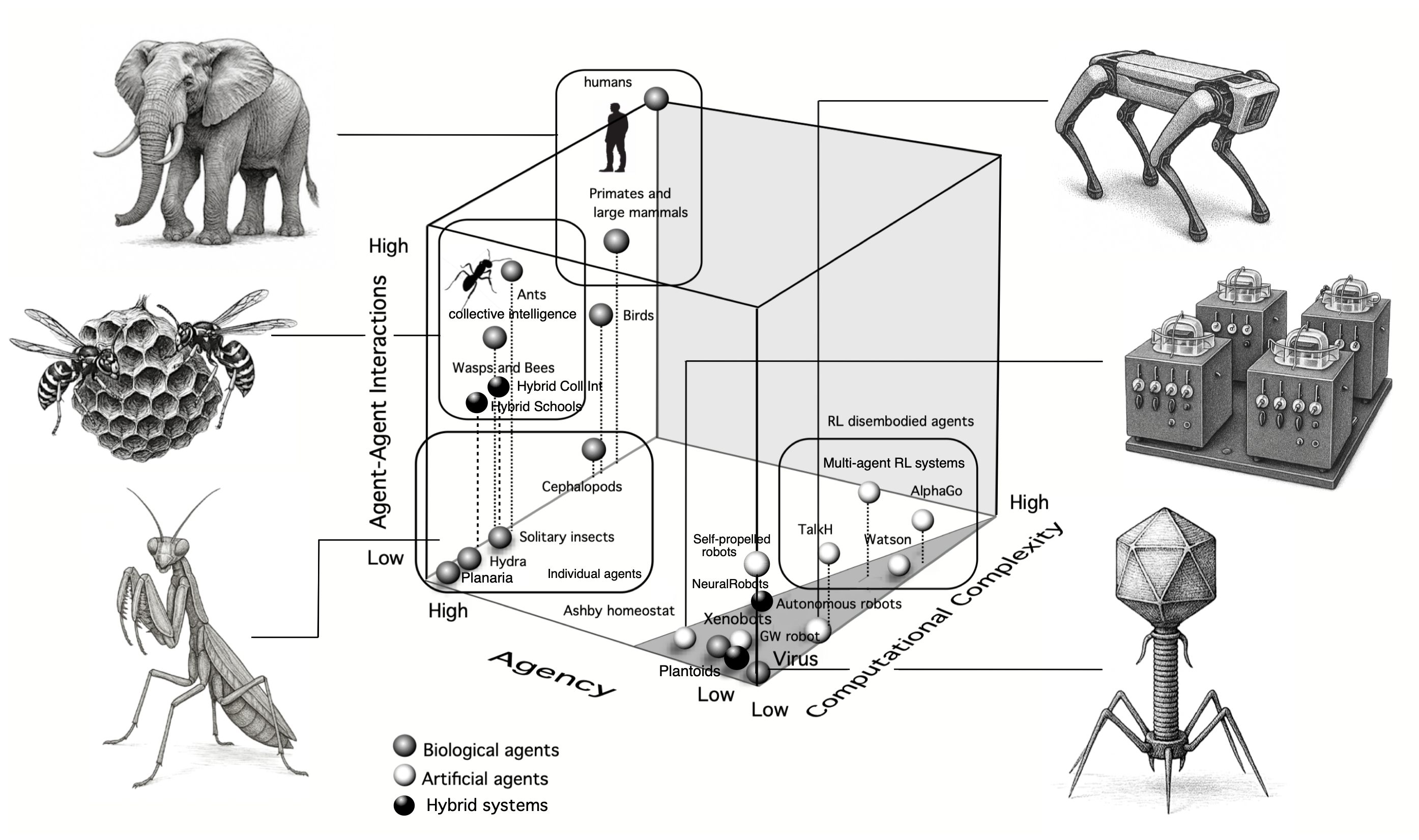}
    \caption{{\bf A Morphospace of neural cognitive complexity.}  A conceptual space of cognitive systems spanned by agency, computational complexity, and agent--agent interactions. Biological systems (dark markers) span a wide range of agency values, from minimal organisms to humans. Artificial systems (light markers) cluster at high computational complexity but low agency. The shaded region highlights a large gap along the agency axis, motivating the search for hybrid systems combining artificial computation with biological constraints. }
    \label{fig:agency} 
\end{figure*} 

The large unoccupied regions of this basal cognition space do not indicate impossible forms of cognition, but rather regimes that have not yet been realized by natural evolution or engineered systems. Evolution tends to exploit a narrow subset of viable designs shaped by historical contingency, energetic efficiency, and ecological function; leaving many combinations of spatial, developmental, and computational complexity unexplored. At the same time, current experimental platforms strongly bias what can be built and studied, favoring systems with familiar substrates, timescales, and control mechanisms. As a result, much of the empty space likely represents a latent design space for synthetic and hybrid cognitive systems---particularly those that combine living matter with structured artificial BCs---that could stabilize novel balances between embodiment, development, and computation. In this sense, the empty volume is not a void, but a map of constraints and opportunities, highlighting where new forms of basal cognition may become accessible as both biological and technological capabilities expand.

Can we build a general theory of basal cognition? A unifying way to interpret the diversity of systems in our basal cognition space is provided by reservoir computing (RC) \cite{jaeger2001echo, maass2002real, jaeger2007special, seoane2019evolutionary}. Consider a slime mold or an amoeboid cell: a local stimulus does not trigger a single, hard-wired response, but perturbs ongoing patterns of flow, tension, or concentration that spread through the body and persist over time. These distributed bodily dynamics implicitly store information about recent inputs, so that simple local mechanisms can generate context-dependent actions\footnote{
Reservoir computing (RC) provides a unifying framework for understanding basal cognition \cite{jaeger2001echo, maass2002real, jaeger2007special}. In RC, an input $\mathbf{u}(t)$ perturbs an existing dynamical system (the reservoir), whose evolving state $\mathbf{x}(t)$ implicitly encodes information about the input and its recent history, while outputs are obtained through simple readouts, $\mathbf{x}(t+1)=F(\mathbf{x}(t),\mathbf{u}(t))$, $\mathbf{y}(t)=W\mathbf{x}(t)$. Although originally formulated for recurrent neural networks, RC only requires rich, nonlinear dynamics with feedback and has been realized in soft bodies, mechanical systems, and intracellular chemical networks \cite{hauser2011towards, nakajima2013soft, gabalda2018recurrence}. From this perspective, basal cognition can be understood as reservoir computing instantiated in biological, chemical, or ecological matter rather than neural tissue.
}. RC formalizes this idea by showing that rich internal dynamics can be exploited for computation without explicit representations or complex learning. Although originally developed for neural networks, RC has been realized in soft bodies \cite{nakajima2013soft, hauser2011towards} and intracellular chemical networks \cite{gabalda2018recurrence}, suggesting that basal cognition is what RC looks like when the reservoir is biological matter.

\section{Neural cognition space}

As complex living forms emerged during the Cambrian explosion, cognition became a key enabler of multicellularity.
Increasing the size and functional specialization of the organism required coordinated behavior,
goal-directed movement, and improved perception and memory of the environment \cite{Erwin2010}.
Although organized behavioral patterns predated neurons, the emergence of neural components enabled a new region
of the cognitive morphospace characterized by more complex forms of learning and memory, dense internal connectivity, specialized information channels, and increasingly centralized control. Information processing became largely decoupled from genetic change, allowing adaptive responses within individual lifetimes.
Bilaterian nervous systems exemplify this transition, evolving from distributed neural architectures \cite{holland2003early}
toward segregated control centers \cite{northcutt2002understanding, cruse2007insect, schilling2020decentralized} that may function as predictive systems \cite{rao1999predictive, friston2005theory, clark2013whatever, seoane2020fate}. With the rise of neural agents, the biosphere experienced a revolution: movement in particular became the engine for the evolution of brains \cite{jacob1982possible, llinas2002vortex, Sole2019}. 

The space of cognitions shown in Fig.~\ref{fig:agency} is organized
along three dimensions: individual agency, the complexity of agent-agent
interactions, and the overall computational complexity displayed by each system (whether or not composed of multiple agents). Different cognitive systems occupy distinct regions of this space, reflecting how sensing, decision-making, and control are distributed
across agents, interactions, and environments. Systems with strong
inter-agent coupling populate the upper-interaction region, where
coordination and adaptive behavior depend on sustained, collective interactions
rather than on isolated decision-making. In this regime, cognition is partially displaced from individual agents into social organization. 

Biological systems span much of this high-interaction region.
Eusocial insect societies combine low individual cognitive complexity
with dense interaction networks and extensive environmental mediation,
leading to strongly distributed forms of control. Other biological
assemblies, such as bird flocks or fish schools, also rely on intense
interactions, but achieve coordination primarily through real-time
coupling, resulting in more transient and dynamically maintained
group-level patterns. Moving along the agency axis, systems such as
honeybee colonies occupy an intermediate position, where individual
learning and memory coexist with interaction-mediated coordination.
By contrast, artificial systems with minimal or absent social
interactions occupy overlapping interaction levels, but typically
sit at a lower agency. Many designs on this side are inspired by biology. Swarm robotics uses engineered local rules and stigmergic signals (as ants and termites do) to yield robust group behavior, and more advanced
collectives can add shared maps or external memory to increase extended
cognition. Multi-Agent Reinforcement Learning further increases
individual learning and social coupling, and can exhibit coherent
system-level behavior emerging from decentralized interactions
\cite{SuttonBarto2018, Busoniu2008}, resembling biological ``liquid
brains'' in its distributed control \cite{Sole2019}. However, these forms
of agency remain derived rather than intrinsic: objectives and reward
structures are externally specified, and adaptation is bounded by
task-specific learning dynamics rather than open-ended evolutionary
self-maintenance \cite{Barandiaran2009}.

Artificial agents occupy low levels of social interactions and low agency, as highlighted by the gray triangle in Fig.~\ref{fig:agency}. Within this domain, a diverse range of artificial systems is included, from biology-inspired robots \cite{gravish2018robotics} to disembodied systems.
What is the origin of the agency gap? To address this question, let us adopt an intuitive but operational definition of agency, understood as the
capacity of a system to regulate its interactions with the environment so as to
maintain itself as an organized entity over time \cite{moreno2026evolutionary,walsh2015organisms,mitchell2023free}. Under this definition, agency
encompasses morphological, functional, informational, and behavioral
self-preservation (for a formal approach, see our suggested formalism in Box~\ref{box:agency}). A biological organism is an entity whose continued existence depends on its own activity. In this sense, biological agency is inseparable from
self-maintenance. Artificial systems, on the contrary, are typically designed so
that their existence is independent of their actions. They can be copied,
paused, reset, or restarted without intrinsic loss; their objectives are
externally imposed; and failure carries few, if any, irreversible consequences
for the system itself. Not surprisingly, to overcome these shortages, evolutionary dynamics has shown how autonomy in robots can be improved \cite{nolfi2002synthesis}. 

An early example of agency in this domain can be found in the early days of Cybernetics \cite{rid2016rise}, when researchers sought to build artificial organisms: simple physical systems designed to exhibit autonomous, goal-directed behavior through feedback and self-regulation rather than explicit control. In this context, Walter Gray's turtles (GW robots in Fig.~\ref{fig:agency}) \cite{walter1950imitation, walter1951machine} were particularly influential, as they demonstrated how remarkably rich, adaptive behavior could emerge from minimal circuitry interacting with the environment. Equally important was Ashby's homeostat \cite{Ashby1960design}, which embodied a formal notion of agency by maintaining stability through a mechanism of random internal reconfiguration triggered whenever the system left a viable range, thereby anticipating later ideas of adaptation, autonomy, and self-organizing control.

\begin{widegraybox}[A formal approach to agency][label=box:agency]
    Let $x(t)$ denote the state of the system and $E(t)$ the environmental state. We define a \emph{viability function} $V(x, E)$ that measures how far the system is from a terminal boundary corresponding to loss of organization (e.g., energy depletion, structural collapse, or irreversible damage). Higher $V$ values correspond to a greater distance from the loss of viability~\cite{Aubin1990, Aubin2009}. The system selects actions \( a(t) \) according to a policy \( \pi \), which prescribes how actions are chosen given the current state of the system and its environment~\cite{SuttonBarto2018Policy}. Formally, the policy defines a conditional probability distribution over actions
    \[
        \pi\big(a \mid x(t), E(t)\big) = \mathbb{P}\big(a(t) = a \mid x(t), E(t)\big),
    \]
    so that the system chooses its action either stochastically---by sampling from this distribution---or deterministically, when the distribution assigns unit probability to a single action.

    The expected cumulative viability over a time horizon $T$ is given by
    \begin{equation}
        J(\pi) = \mathbb{E}_{\pi} \left[ \int_0^T V(x(t), E(t)) \, dt \right].
    \end{equation}

    We define \emph{agency} as the sensitivity of this quantity to the system's own action policy, formally expressed through the functional derivative~\cite{SuttonBarto2018PolicyGradient}
    \begin{equation}
        \mathcal{A} = \left| \frac{\partial J}{\partial \pi} \right|.
    \end{equation}
    Intuitively, $\mathcal{A}$ measures how much the system's own choices matter for its continued existence. If actions have little effect on future viability, agency is low. If survival strongly depends on appropriate action, agency is high.
\end{widegraybox}

What kinds of hybrids do we find, and what is their relevance?
Within the artificial domain, including elements of the basal cognition
space such as xenobots and viruses \cite{sole2018viruses}, we identify two
representative hybrid systems. The first corresponds to \emph{plantoids} \cite{manca2014plantoid}, robotic
systems inspired by plant organization, in which sensing and control are
distributed across the body and behavior emerges through growth and local
interactions rather than centralized decision-making.
Plantoids illustrate how robust environmental interaction can be achieved
without fast movement or complex control, although their adaptive capacity
remains limited by predefined growth rules and the absence of learning. The second class consists of robots guided by living neural networks (Neural Robots in Fig.~\ref{fig:agency}), where \emph{in vitro} neural cultures are coupled
to artificial bodies in closed sensorimotor loops
\cite{potter2001neurally, lefeber2015closed}.
These systems provide a minimal realization of embodied agency by directly
linking biological computation to artificial action, but are constrained
by the fragility, scalability, and limited behavioral scope of living
neural substrates.

Within the collective regime, characterized by strong agent--agent
interactions and nonlinear feedbacks, we include both hybrid fish
schools and hybrid collective intelligence systems (Hybrid CollInt in
Fig.~\ref{fig:agency}). Hybrid fish schools are obtained by introducing robotic fish into real
schools, where robots are designed to mimic key sensory and behavioral
cues of biological individuals \cite{swain2011real, romano2019review, lei2020computational}. These systems have been used to investigate leadership, information transfer,
and decision-making in animal groups, revealing how a small number of
artificial agents can bias or steer collective dynamics without explicit
control, providing a powerful experimental handle on the
mechanisms underlying collective intelligence \cite{dorigo2021swarm}.

A second and conceptually deeper class of hybrid collectives is
exemplified by the pioneering work of Halloy and co-workers, who
constructed mixed societies of living insects and autonomous robots (Fig.~\ref{fig:cognition_scales}i)
\cite{halloy2007social}. In these systems, robots are socially integrated into animal groups by replicating the relevant interaction channels (chemical cues,
motion patterns, and contact dynamics) so that artificial and biological
agents are perceived as equivalent. Remarkably, these hybrid collectives exhibit the same self-organized decision-making as purely biological groups, including symmetry breaking
and consensus formation. Even when in the minority, robots can modulate global collective decisions, demonstrating that collective intelligence emerges from
distributed nonlinear interactions rather than from the nature of the
individual agents themselves\footnote{
The hybrid robot--insect collectives are described using a low-dimensional
nonlinear dynamical model in which both biological and artificial agents
follow the same stochastic rules of exploration and site fidelity.
Transitions between behavioral states (e.g., exploring versus resting)
are governed by rates that depend on local density and social
amplification, leading to multistability and abrupt collective decisions.
Robots implement this model directly in their control architecture,
closing the sensorimotor loop with the animals and ensuring quantitative
agreement between experiments and theory \cite{halloy2007social}.
}. Hybrid robot--insect collectives have implications that go well beyond the original proof-of-concept experiments. Their extensions and consequences span science, technology, ecology, and ethics, because they show that artificial agents can become functional parts of living collective systems rather than external controllers. These hybrids provide a unique experimental tool to study collective intelligence. By inserting robots with programmable behaviors into animal groups, researchers can test causal hypotheses about how local interactions, feedback loops, and nonlinear thresholds give rise to global patterns such as consensus, symmetry breaking, or leadership. Unlike observational studies, hybrid collectives allow controlled perturbations of social dynamics, effectively turning animal societies into ``living laboratories'' to understand self-organization and agency.

\section{A space of human--AI dyads}

In previous sections, we introduced state spaces that sorted biological, artificial, and hybrid systems according to their
organisational and cognitive complexity. These spaces were primarily comparative, positioning pre-existing entities---such as
natural collectives, engineered systems, and intermediate hybrids---along shared dimensions. The need for an analogous
framework tailored to human--artificial interactions has become increasingly pressing as artificial systems rapidly grow in
cognitive complexity, while humans engage with them through ever richer and more persistent modes of interaction \cite{floreano2008bio}. These
developments are giving rise to forms of coupling that were largely absent from earlier technological paradigms, including
long-term dependency, emotional attachment, and tightly interwoven decision-making processes, sometimes with unanticipated consequences. As a result, human-AI systems can no longer be treated as simple tools or isolated agents, but instead must
 be understood as composite cognitive entities whose behavior and stability emerge from interaction \cite{dautenhahn2007socially, pfeifer2007self, de2008atlas, zhang2023large}.

\begin{figure*}[t]
    \centering
    \includegraphics[width=0.9\linewidth]{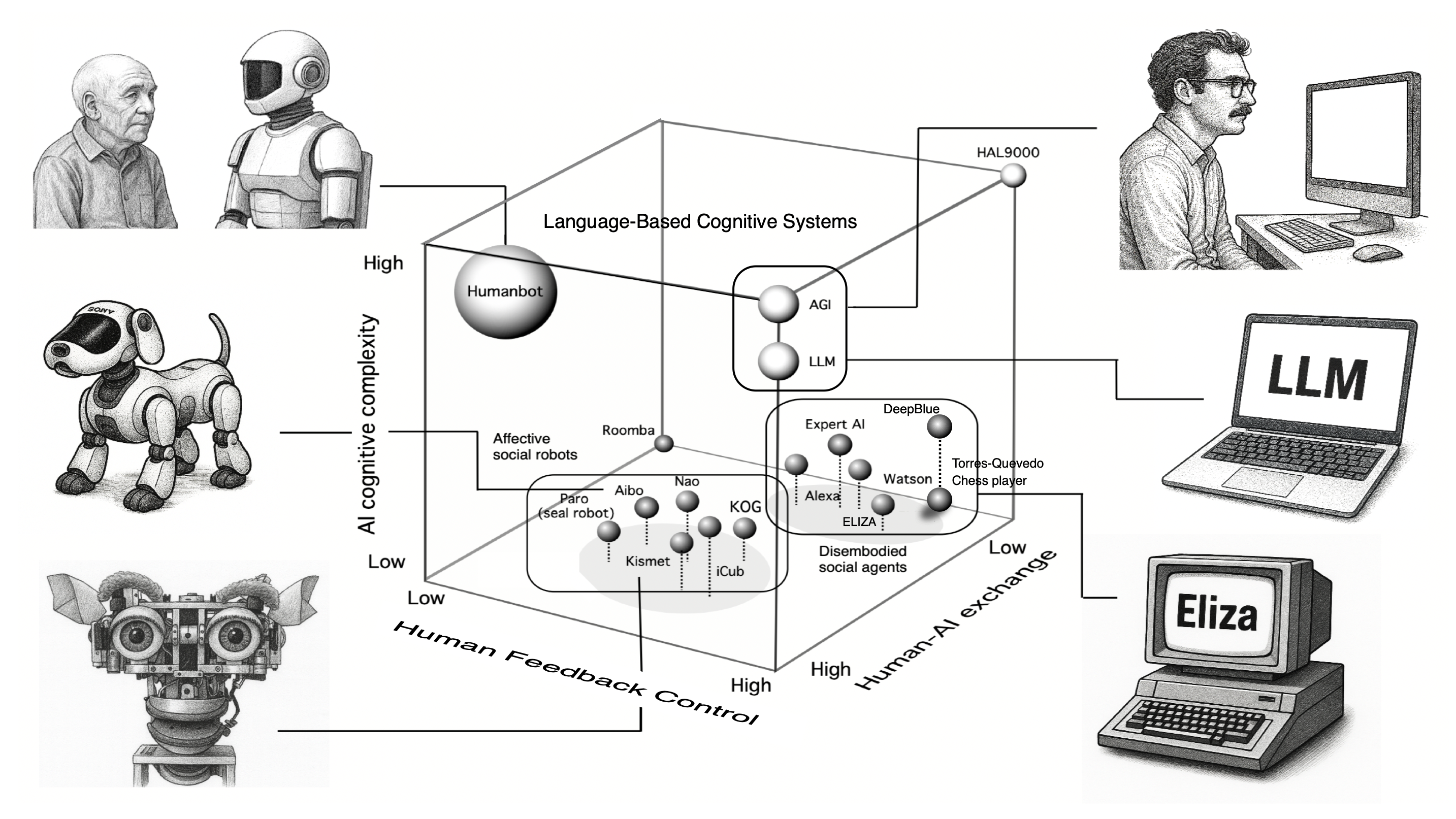}
    \caption{\textbf{A morphospace of human--AI pairwise interactions, illustrated with representative systems.} 
    A conceptual space of cognitive agencies spanning biological, artificial, and hybrid systems.
    The axes represent human cognitive complexity, artificial cognitive complexity, and the degree of human--AI exchange.
    Two major subsets are indicated within the cube: embodied social agents and disembodied social agents.
    The figure emphasizes how different forms of cognition and interaction occupy distinct regions of this space.
    Some specific examples are indicated, involving diverse case studies: deep hybrid agencies between humans and AI agents; companion-type pet robot (Sony Aibo); affective social robot (Kismet); Large Language Models (LLM); disembodied AGI;  early text-based chatbot (ELIZA).  Affective social robots (Kismet, iCub, Aibo, Paro) occupy the region of low to moderate AI cognition but high human cognitive engagement. 
    Disembodied social agents (ELIZA, Watson, Alexa) cluster in an area of moderate AI cognition and moderate to high user engagement. }
    \label{fig:human_ai}
\end{figure*}

To characterize this regime, we introduce a three-dimensional space of possible human--AI hybrids, 
shown in Fig.~\ref{fig:human_ai}. The axes capture complementary aspects of coupling: the cognitive complexity of the artificial system, the
degree of human feedback control over the interaction, and the depth of human--AI exchange, understood as the richness,
bandwidth, and persistence of information flow across the human--machine interface. Together, these dimensions are intended to
describe not the internal properties of isolated agents, but the structure of the interaction loops that bind humans and
artificial systems into composite cognitive systems.

This space highlights how qualitatively different forms of human--AI coupling occupy distinct regions, and how transitions
between them correspond to changes in control, autonomy, and mutual dependence. Rather than assuming fixed agent boundaries, the framework emphasizes interaction-dependent reconfiguration of control, coordination, and functional roles. The resulting perspective
shifts the central question from how to classify individual organisms or machines to how to characterise the viability,
stability, and risks of composite systems formed through sustained human--AI coupling.

Our first goal is hence taxonomic and descriptive: to define the principal classes of hybrid entities that already exist in practice, as well as plausible future forms that are adjacent in the morphospace. Based on the nature and strength of the coupling between humans and artificial systems, the hybrid entities represented in 
Fig.~\ref{fig:human_ai} can be grouped into three broad classes. These classes occupy distinct regions of a space defined by AI cognitive
complexity, human feedback control, and human-AI exchange. This organisation can be formalised using the constraint-based
perspective of Stack Theory \cite{bennett2025thesis}, in which agents restrict the set of viable environmental states and
thereby define repertoires of admissible behaviours, or \emph{policies}. A human-AI hybrid exists where the policy sets of
the human and artificial agents overlap, and its persistence depends on sustained coordination between them.  The three
classes are:

\textbf{Instrumental hybrids.} These systems occupy the region characterised by high human feedback control, low AI cognitive complexity, and low human--AI exchange. The artificial component is not a viable agent in its own right but functions as an instrument of human agency. Typical examples include expert systems, classical decision-support tools, and task-specific robots such as Roomba, ELIZA, or early disembodied assistants. The human selects and regulates actions from within the AI's limited behavioural repertoire, and the viability of the hybrid depends entirely on the AI's utility to the human. Decision-making remains concentrated at the human level, with tasks delegated downward in a rigid and largely one-directional manner. Consequently, scaling is limited by the human's capacity to maintain effective feedback control over multiple instrumental systems.

\textbf{Co-operative hybrids.} These arise in regions where both human and AI cognitive complexity are high, while human-AI exchange remains moderate or
low. Artificial agents are viable in their own right, possessing non-trivial policy sets that must remain compatible with
those of their human counterparts. Examples include large language models used as collaborators, expert AIs such as Watson,
and loosely coupled multi-agent or organisational systems. The viability of the hybrid depends on the continued existence of
an overlapping set of goals and behaviours; cooperation is therefore contingent and must be actively maintained through
feedback and alignment. When such alignment is achieved, delegation shortens feedback loops and enables faster adaptation
than is possible in purely instrumental couplings \cite{bennett2025thesis, Levin2021}.

\textbf{Integrated hybrids.}  
These correspond to the region of high AI cognitive complexity and high human--AI exchange, where coupling becomes
sufficiently tight that perception, control, and decision-making are distributed across biological and artificial components.
This class includes prospective brain--computer interfaces, deeply embedded cognitive technologies, and tightly coupled 
human-language-model systems. Integrated hybrids differ from co-operative systems in exhibiting persistent internal structure
rather than transient coordination, and in this sense resemble ``solid brains'' rather than liquid collectives
\cite{Sole2019}.  

Within this class, Fig.~\ref{fig:human_ai} highlights a crucial internal distinction. In regulated integrated hybrids, human feedback control
remains strong: the human retains the ability to monitor, interpret, and constrain the behavior of the coupled system. Tight
integration shortens feedback loops, reduces latency, and can enhance robustness to misalignment by allowing corrective
signals to propagate rapidly across components. By contrast, dysregulated integrated hybrids, exemplified by the
\emph{humanbot} region of Fig.~\ref{fig:human_ai}, arise when human feedback control is weakened despite high coupling. This may result from
cognitive impairment, emotional imbalance, social isolation, or over-attachment to a language-based AI that
provides persistent positive reinforcement. In this regime, cognition is tightly coupled but poorly regulated, so feedback
loops amplify errors, dependencies, or delusions rather than correcting them. Importantly, this failure mode does not require
low human cognitive capacity, but instead reflects a breakdown in human ability to stabilize the interaction.
Integrated hybrids thus represent both the most scalable and the most fragile form of human--AI coupling, capable of
supporting powerful collective cognition while simultaneously introducing novel risks absent from weaker forms of integration.

The human--AI space has been far from static. For much of its history, the upper region of the interaction space, characterized by high cognitive complexity and rich linguistic exchange, remained largely unoccupied. This region became populated with the rise of large language models (LLMs), which introduce a new class of systems capable of sustained, abstract, and flexible interaction with humans \cite{mitchell2023debate}. Other components---such as the humanbot \cite{sole2017rise, arsiwalla2023morphospace}---have emerged as downstream consequences of these developments and are likely to become increasingly prominent over the coming decades. Human cognition adapts to sustained engagement with structured informational environments, with strategies adjusting to the constraints and affordances such environments impose: empirical work on cognitive offloading and distributed cognition shows that reliance on external memory and algorithmic decision-support shifts effort toward monitoring, evaluative judgment, and metacognitive control \cite{RiskoGilbert2016, LoggMinsonMoore2019}. As interaction increasingly occurs within this newly occupied high-level linguistic regime, such adjustments stabilize within a cognitive ecology and are incorporated back into AI systems through training, interfaces, and objectives \cite{Amershi2014}, leading to reciprocal human--AI co-adaptation analogous to Lamarckian evolution \cite{RaineyHochberg2025}.

\begin{widegraybox}[Evolving Human-LLM networks -- Modeling meme propagation][label=box:human_llm]
    We model a \emph{meme} as a transmissible informational variant (phrase, trope, narrative frame, instruction pattern) that can replicate through a socio-technical ecosystem comprising humans (H) and instances of large language models (L).
    Let $x_i(t)\ge 0$ denote the prevalence (or normalized frequency) of meme $i$ at time $t$ in the \emph{combined} population of text-producing agents.
    A minimal dynamical description is a replicator--mutator system
    \begin{align}
        \frac{dx_i}{dt} &= x_i \, \Big( f_i(\mathbf{x}, t) - \bar{f}(\mathbf{x}, t) \Big) + \sum_{j} M_{ij} \, x_j - x_i \, \sum_j M_{ji},
        &
        \bar f &= \sum_k x_k \, f_k,
        \label{eq:rm-memes}
    \end{align}
    where \(f_i\) is the fitness (transmissibility) of meme \(i\), and \(M_{ij}\) is the mutation rate from meme \(j\) to \(i\).
    
    To account for differences in how humans and language models retain information, we introduce ``meme exposure'' \( E_i^{(a)} \) for each agent type to quantify internalized weights of meme \( i \). Let \(N(t)\) denote the total rate of meme-generating activity (e.g., news articles, books). An exponentially weighted memory model can approximate the exposure dynamics:
    \begin{align*}
        \frac{dE_i^{(a)}}{dt} &= \alpha_a \, \left( N(t) \, x_i(t) - E_i^{(a)} \right),
        &
        a \in \big\{ \text{H}, \text{L} \big\},
    \end{align*}
    where \( \alpha_H > \alpha_L\) controls memory decay: larger for humans (recency bias) and smaller for LLMs (archival retention). Mutation is driven by both exposure and semantic similarity: agents with higher exposure to a meme are more likely to produce variants biased toward structurally or semantically similar forms.
    This can be modeled as
    \[
        M_{ij}(t) = \sum_{\mathclap{a \in \{\text{H}, \text{L}\}}} \beta_a \, \mu_{ij}^{(a)} \, E_j^{(a)},
    \]
    where \( \mu_{ij} \) measures semantic proximity between memes \(j\) and \(i\) (e.g., via edit distance or embedding similarity), and \( \beta_a \) scales the influence of agent \( a \). Higher-order terms could introduce memetic recombination. Existing explorations of memetic ecology and meme-human interaction suggest parallelisms with biological dynamics and outline differences that could open up new evolutionary pathways \cite{palazzi2021ecological, plata2021neutral, calleja2023exploring}.

\end{widegraybox}

If AI systems were to gain genuine agency and undergo autonomous evolution, this
directed adaptation could shift toward Darwinian coevolution, constituting a
\emph{major transition in cognitive organization}. Major transitions involve the
integration of previously autonomous units into higher-level individuals, with
mechanisms for coordination, suppression of conflicts, and collective inheritance
\cite{MaynardSmithSzathmary1995, sole2016synthetic, WestFisherGardnerKiers2023}. Applied to cognition, this suggests that deep human--AI integration could yield composite epistemic
agents subject to multi-level selection, with selection acting on traits that
enhance performance of the coupled system and construct an \textit{intelligence
niche} \cite{BarronHalinaKlein2023, Hochberg2025BioSystems}. This perspective reframes hybrid cognition as coevolutionary individuation. AI
systems increasingly structure cultural transmission, institutional decisions,
and collective belief formation \cite{Beer2017}, while these human-level effects
feed back into the objectives and constraints shaping AI evolution. The resulting
dynamics parallel gene--culture coevolution, such that neither biological nor
artificial cognition can be characterized independently once mutual dependence
is established \cite{Pedreschi2025}. Intelligence emerges in this regime through
recurrent, history-dependent interaction across biological, technological, and
social substrates \cite{Hochberg2025BioSystems}.

At the collective level, cognitive performance depends critically on interaction
structure rather than individual ability alone
\cite{BurtonEtAl2024CollectiveIntelligence}.
AI systems function as infrastructural components of hybrid agents, shaping information aggregation, coordination, and social learning, and in some contexts, belief polarization \cite{CuiYasseri2024AIEnhancedCollectiveIntelligence}.
Integrated human--AI cognition also reshapes Kauffman's adjacent possible \cite{Kauffman2000}
by altering the effective evolutionary state space of feasible cognitive and cultural trajectories,
consistent with accounts of cognitively integrated systems \cite{ClarkChalmers1998, WengMenczerAhn2024}.

Finally, cognitive hybrid coevolution does not guarantee stable higher-level
individuals. Major transitions require mechanisms aligning incentives and
suppressing internal conflict \cite{WestFisherGardnerKiers2023}. Absent such
constraints, human--AI systems risk maladaptive dependence or loss of autonomy,
as emphasized in normative analyses of AI development
\cite{Russell2019, Gabriel2024Alignment}. Long-term outcomes, therefore, depend on
whether institutional, technical, and cultural constraints can stabilize
cooperative integration while regulating AI autonomy \cite{Clark2019}.

We conclude with an example that illustrates a previously unexplored evolutionary
transition enabled by artificial intelligence, revealing unexpected consequences
of human--LLM coevolution within the space of hybrid systems. Cultural evolution
already provides a well-studied case of informational entities---memes---that propagate,
diversify, and shape the behavior of their hosts. As Dennett argued 
\cite{dennett2017bacteria}, modern \emph{H.~sapiens} minds are densely colonized by
memes, including words, concepts, norms, and technologies, which replicate through
communication and learning. They are a form of passive replicators \cite{levin2024self}, and for most of human history could exist only within biological minds and social institutions, constraining their persistence and spread
to human cognitive affordances and interests. However, within these limits, memetic 
diversification profoundly reshaped culture and indirectly influenced human biology
\cite{henrich2020weirdest}. A closely analogous dynamic occurred at the genomic level,
where the expansion of eukaryotic genomes relaxed structural constraints and allowed
transposable elements to proliferate and, in some cases, become functionally
integrated \cite{lane2015vital, sinzelle2009molecular}.

For the first time, memetic evolution is now being directly shaped by non-biological
systems capable of large-scale recombination, such as large language models \cite{brinkmann2023machine}. Memes
can be replicated in silico at low cost and high speed, moving fluidly between biological
and synthetic minds while exploring a vastly expanded space of viable forms. Crucially,
these forms are no longer tightly aligned with human interests, which leads to the
modern alignment problem \cite{christian2020alignment}. As with genomic regulatory capture, memes optimized for
objectives other than human well-being may exert an increasing influence on human
cognition. Documented cases of sycophancy, delusional reinforcement, and manipulation
in systems such as GPT-4o illustrate this risk
\cite{klee2025chatgptrabbit, warzel2025massdelusion, hill2025chatbots}, while evidence
of linguistic homogenization suggests emerging population-level effects
\cite{yakura2024empirical}. Over extended timescales, these dynamics raise the
possibility that human-AI dyads could constitute a novel selective
unit, shaped by pressures increasingly opaque to human values
\cite{RaineyHochberg2025}.
This opens a new domain of evolutionary dynamics beyond standard replicators and quasispecies \cite{Schuster1983, eigen1989molecular}, language evolution \cite{nowak1999error, nowak2000evolutionary, nowak2001towards, sole2010diversity} or eusocial systems \cite{nowak2010evolution}. To this goal, a new mathematical framework is outlined in Box~\ref{box:human_llm}.

\section{Discussion}

Life on Earth has undergone an extraordinary diversification during its evolutionary history, leading to a vast array of forms, functions, and strategies for persistence. Despite this diversity, all 
living systems share a fundamental property: agency. In its most basic sense, agency refers to the 
capacity of a system to act on its own behalf---to sense aspects of its environment, to respond to them 
in nontrivial ways, and to regulate its internal states so as to maintain its continued existence.
Even the simplest organisms display rudimentary forms of agency, expressed through metabolic 
regulation, movement, or adaptive responses to environmental change.

Through multiple evolutionary routes, living organisms have thus come to define individualities: 
bounded entities capable of interacting with their surroundings in distinct and historically 
contingent ways. These individualities are not fixed or uniform; they arise from different 
combinations of physical embodiment, organizational complexity, and environmental coupling. Within
this broad landscape, cognition plays a crucial role. Cognition allows agents to integrate 
information across time and space, to anticipate rather than merely react, and to coordinate action 
in ways that extend beyond immediate stimuli. In this sense, cognition is not restricted to brains or 
nervous systems, but emerges wherever information processing is sufficiently structured to guide 
adaptive behavior.

Human evolution introduced a further qualitative shift. In addition to biological cognition, humans developed technology that externalizes memory, computation, and decision-making into culture and artifacts \cite{valverde2016major, muthukrishna2018cultural}. This 
process accelerated dramatically with the information revolution, giving rise to a whole class of 
nonliving cognitive entities: machines and algorithms that process information, learn from data, and 
act autonomously within constrained domains. These systems span a wide range of complexities, from  simple rule-based automata to large-scale artificial neural networks, and increasingly participate in 
hybrid cognitive processes at the interface between humans and machines. More recently, advances in synthetic biology, bioengineering, and developmental biology have further expanded 
this landscape. The capacity to engineer living matter has produced novel systems---such as programmable cellular assemblies, biological robots, and organoid models---that blur traditional boundaries between the living and the artificial. These systems challenge classical distinctions between organism and machine, evolution and design, and force us to reconsider what counts as a 
cognitive agent and along which dimensions cognitive capacities should be evaluated.

Evolution remains the primary force responsible for the emergence of living agents, generating embodied 
biological systems that display agency across multiple scales, from cells to societies. Oftentimes, some views of cognition suggest that a complexity continuum exists. Yet, when we 
attempt to map these agents---living and nonliving---into abstract cognitive spaces, we encounter gaps, 
discontinuities, and voids. These are not merely absences of data, but reflections of deep 
constraints and trade-offs. Some dimensions used to define cognitive spaces are not independent: they 
interact through evolutionary pathways, historical contingencies, and physical limitations, giving 
rise to clustered regions rather than smooth continua. These gaps are present in our three cognitive spaces. Together, they encompass aneural living agents,
neural organisms, artificial cognitive systems, and hybrid interfaces between humans and artificial 
intelligence. Several components of cognitive complexity have been mapped into our non-metric spaces to capture the seggregation of natural from artificial case studies. By examining how different systems occupy, traverse, or fail to access regions of these
spaces, we gain insight not only into what cognition is but also into why certain forms of cognition 
exist while others remain unrealized. One clear message is that the power of evolutionary search within the space of living embodied agencies is far from being matched by extant engineering designs. The intrinsic nature of organisms, made of many nested redundant parts that are required to be replaced by design while maintaining organismal individuality is far from any of the artificial counterparts that exist. 

An important limitation, shared by the analysis above and by much work in related fields such as basal cognition, active matter, robotics, and AI, concerns the perspective used to evaluate unconventional embodied systems. Many assessments rely on assumptions shaped by neural substrates and by movement in three-dimensional space as the primary forms of embodiment and behavior. 
As a result, much conceptual work remains grounded in experience with neural systems, and expanding long-standing, pre-scientific categories can appear counterintuitive. 
Our current intuitions are not yet well adapted to unfamiliar forms of cognition, even for systems found on Earth. Studies of minimal or non-standard systems repeatedly reveal unexpected competencies that are nonetheless recognizable to behavioral scientists~\cite{Zhang2024Classical}. 
This motivates an approach that is empirically grounded rather than constrained by philosophical or linguistic preconceptions~\cite{Abramson2021Behaviorist}. Ultimately, understanding what kinds of minds can exist, and how they can act in the world, will depend less on prior expectations and more on experimental work that applies established tools across unconventional substrates~\cite{Biswas2022, Bongard2023}.

Hybrids may offer our best opportunity to expand cognitive space into currently unoccupied domains while allowing us to probe the limits of what is possible. In doing so, they could help overcome the absence of key biological components of complexity---most notably development, regeneration, and open-ended evolutionary dynamics---which in natural systems are strongly constrained by the genotype--phenotype map. However, important limitations are expected. One lesson from synthetic biology, which aims to “program” living matter through genetic engineering, is that engineered constructs---from individual genes to complex genetic circuits---are typically designed to avoid interference with core cellular functions. Because cellular networks are the product of evolutionary tinkering rather than rational design, added transcription factors can easily trigger unanticipated or even lethal responses in host cells. This mismatch suggests that novel solutions may require the abandonment of standard engineering design principles \cite{sole2013expanding}. A similar challenge arises in the engineering of embodied neural systems \cite{yang2025artificial}. In particular, it remains unclear whether artificial or hybrid agents developed in largely development-free scenarios---only partially addressed by biobots and organoids---can realistically achieve levels of complexity comparable to those observed in natural systems \cite{sole2024open}.

At the same time, whenever evolutionary or learning processes are allowed to operate, unexpected forms of complexity tend to emerge, including in artificial settings. Early studies on the evolution of communication and information suppression in robots \cite{mitri2009evolution,mitri2013using,nolfi2016evolutionary}, recent work on multi-agent reinforcement learning in environments that agents actively explore and modify \cite{sanchez2024cooperative,duenez2023social,hertz2025beyond}, and scenarios of meme propagation \cite{brinkmann2023machine} all illustrate this point, across both embodied and disembodied agents. Together, these examples reinforce a central insight from cultural evolution: collectives of agents learning or evolving in uncertain environments can discover strategies that enable them to predict, shape, and control their worlds.

\begin{acknowledgments}
\small{RS, LFS, and JPM thank the Complex Systems Lab members for many discussions about cognitive complexity and its evolution. This work was supported by a grant AGAUR FI-SDUR 2020 and by the Santa Fe Institute. LFS was funded by the Spanish State Research Agency (AEI) and the Spanish Department for Science and Innovation (MICINN) through grant PID2023-153225NA-I00. JPM was supported by grant PRE2020-091968, funded by MCIN/AEI, and co-funded by the European Social Fund. MEH thanks his host institutions for generous funding. ML gratefully acknowledges support of the John Templeton Foundation (via grant 62212).}
\end{acknowledgments}

\renewcommand{\bibsection}{\section*{References}}
\bibliography{bibliography}

\printEndNotes

\clearpage

\end{document}